\documentclass[aps,prl,twocolumn,showpacs,nosuperscriptaddress,groupedaddress]{revtex4}  % for review and submission
\usepackage{graphicx}  % needed for figures
\usepackage{bm}        % for math
\usepackage{amssymb}   % for math
\usepackage{amsmath}   % for line breaking
\begin{document}

\title{Evolution arrests invasions of cooperative populations} 
\author{Kirill S. Korolev}
\affiliation{Department of Physics and Graduate Program in Bioinformatics, Boston University, Boston, Massachusetts 02215, USA}
\email{korolev@bu.edu}

\begin{abstract}
Population expansions trigger many biomedical and ecological transitions, from tumor growth to invasions of non-native species. Although population spreading often selects for more invasive phenotypes, we show that this outcome is far from inevitable. In cooperative populations, mutations reducing dispersal have a competitive advantage. Such mutations then steadily accumulate at the expansion front bringing invasion to a halt. Our findings are a rare example of evolution driving the population into an unfavorable state and could lead to new strategies to combat unwelcome invaders. In addition, we obtain an exact analytical expression for the fitness advantage of mutants with different dispersal rates.
\end{abstract}

\maketitle 

%*************************************************************************
% Introduction
%*************************************************************************

Locust swarms, cancer metastasis, and epidemics are some feared examples of spatial invasions. Spatial spreading is the only mechanism for species to become highly abundant, whether we are considering a bacterial colony growing on a petri dish~\cite{korolev:sectors,wakita:expansion} or the human expansion across the globe~\cite{templeton:africa}. Many invasions are unwelcome because they threaten biodiversity~\cite{hooper:biodiversity}, agriculture~\cite{gray:worm_expansion}, or human health~\cite{brockmann:networks}. Unfortunately, efforts to control or slow down invaders often fail in part because they become more invasive over time~\cite{thomas:acceleration}. The evolution of invasive traits and invasion acceleration has been repeatedly observed in nature from the takeover of Australia by cane toads~\cite{phillips:toad_acceleration} to the progression of human cancers~\cite{merlo:review,korolev:perspective}.

Selection for faster dispersal makes sense because it increases the rate of invasion and allows early colonizers to access new territories with untapped resources. A large body of theoretical~\cite{shine:spatial_sorting, korolev:wave_splitting, benichou:acceleration} and experimental work~\cite{thomas:acceleration, phillips:toad_acceleration, ditmarsch:swarming} supports this intuition in populations that grow non-cooperatively, i.e., when a very small number of organisms is sufficient to establish a viable population. Many populations including cancer tumors~\cite{berec:multiple_allee,cluttonbrock:mongoose,merlo:review,korolev:perspective,cleary:cancer_cooperation}, however, do grow cooperatively, a phenomenon known as an Allee effect in ecology~\cite{courchamp:allee_review}. In fact, cooperatively growing populations can even become extinct when the population density falls below a critical value, termed the Allee threshold~\cite{scheffer:review, dai:science}. We find that the intuitive picture of ``the survival of the fastest'' fails for such populations, and natural selection can in fact favor mutants with lower dispersal rates. Over time, repeated selection for lower dispersal leads to a complete arrest of the spatial invasion.

To understand when invasions accelerate and when they come to a halt, we analyzed a commonly used mathematical model for population dynamics that can be tuned from non-cooperative to cooperative growth by changing a single parameter. We considered the competition between two genotypes with different dispersal abilities and computed their relative fitness analytically. Our main result is that selection favors slower dispersal for a substantial region of the parameter space where the Allee threshold is sufficiently high. Numerical simulations confirmed that evolution in such populations gradually reduces dispersal and eventually stops the invasion even when multiple mutants could compete simultaneously and other model assumptions were relaxed.    

%*************************************************************************
% Mathematical Model
%*************************************************************************
Selective pressure on the dispersal rate can be understood most readily from the competition of two types~(mutants, strains, or species) with different dispersal abilities as they invade new territory. For simplicity, we focus on short-range dispersal that can be described by effective diffusion and only consider the dynamics in the direction of spreading. Mathematically, the model is expressed as

\begin{equation}
	\begin{aligned}
		& \frac{\partial c_{1} }{\partial t} = D_{1}\frac{\partial^2 c_{1}}{\partial x^2} + c_{1}g(c),\\
		& \frac{\partial c_{2} }{\partial t} = D_{2}\frac{\partial^2 c_{2}}{\partial x^2} + c_{2}g(c),
	\end{aligned}
	\label{eq:model_main}
\end{equation}

\noindent where~$c_{1}$ and~$c_{2}$ are the population densities of the two types that depend on time~$t$ and spatial position~$x$; $D_{1}$~and~$D_{2}$ are their dispersal rates; and~$g(c)$ is the density-dependent per capita growth rate. We assume that~$g(c)$ is the same for the two types and depends only on the total population density~$c=c_1+c_2$. Since slower-dispersing types often grow faster because of the commonly observed trade-off between dispersal and growth, our results put a lower bound on the fitness advantage of the type dispersing more slowly. In the Supplemental Material, our analysis is further generalized to account for the different growth rates of the types~\cite{supplement}.

For~$g(c)$, we assume the following functional form, which has been extensively used in the literature~\cite{korolev:wave_splitting, aronson:allee_wave, fife:allee_wave, courchamp:allee_review} because it allows one to easily tune the degree of cooperation in population dynamics from purely competitive to highly cooperative growth:

\begin{equation}
	g(c)=r(K-c)(c-c^{*})/K^{2}. 
	\label{eq:growth}
\end{equation}

\noindent Here,~$r$ sets the time scale of growth,~$K$ is the carrying capacity, i.e. the maximal population density that can be sustained by the habitat, and~$c^*$ is a parameter that determines the degree of cooperation and is known as the Allee threshold. For~$c^{*}<-K$, the types grow non-cooperatively because~$g(c)$ monotonically decreases from its maximal value at low population densities to zero when the population is at the carrying capacity and interspecific competition prevents further growth. Population grows cooperatively for higher values of~$c^{*}$ because the per capita growth rate reaches a maximum at nonzero density that strikes the balance between interspecific competition and facilitation. For~$c^{*}>0$, the effects of cooperative growth become particularly pronounced. Indeed, the growth rate is negative for~$c<c^*$ and, therefore, small populations are not viable. Such dynamics, known as the strong Allee effect, arise because a critical number of individuals is necessary for a sufficient level of cooperation~\cite{courchamp:allee_review,korolev:perspective}.

%*************************************************************************
% Results
%*************************************************************************

\begin{figure}
	\includegraphics[width=\columnwidth]{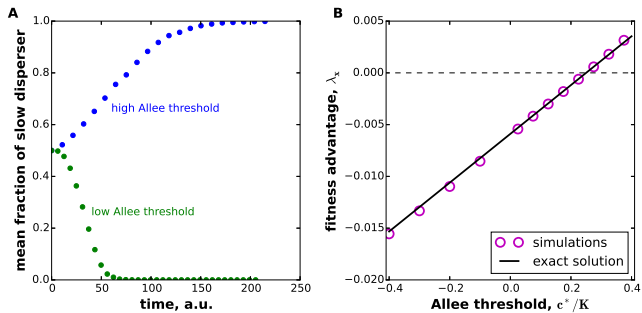}
\caption{The effects of cooperative growth on the evolution of dispersal during invasion. (A) Simulations of the competition between a slow~($D_1=0.5$) and a fast~($D_2=1$) disperser during a spatial expansion. The fraction of the slower disperser decreases in populations with a low Allee threshold~($c^*=0.2$), but increases in populations with high Allee threshold~($c^*=0.35$). (B) The fitness advantage of the slower disperser~($D_1/D_2=0.95$) changes from negative~(deleterious) to positive~(beneficial) as the Allee threshold is increased. In simulations, we never observed the coexistence of the two types; instead extinction is observed for the types that are deleterious when rare~($\lambda<0$), and complete fixation is observed for the types that are beneficial when rare~($\lambda>0$).}
\label{fig:switch}
\end{figure}

We first tested whether unequal dispersal rates lead to fitness differences between the two types by solving Eq.~(\ref{eq:model_main}) numerically~(see Supplemental Material). When population growth was non-cooperative, we found that the faster-dispersing species have a competitive advantage in agreement with the current theory~\cite{thomas:acceleration, shine:spatial_sorting, benichou:acceleration}. Quite unexpectedly, the opposite outcome was observed for strongly cooperative growth: The type with the lower dispersal rate became dominant at the expansion front and eventually took over the population~(Fig.~\ref{fig:switch}a)! 

To understand this counterintuitive dynamics, we examined how the relative fitness of the two types depends on the magnitude of the Allee threshold~$c^*$. In the context of spatial expansions, there are two complementary ways to quantify the fitness advantage of a mutant. The first measure~$\lambda$ is the exponential growth rate of the mutant similar to what is commonly done for populations that are not expanding; a negative~$\lambda$ corresponds to decay not growth. The second measure~$\lambda_{x}$ is the growth rate of the mutant not in units of time, but rather in units of distance traveled by the expansion. The two measures are related by~$\lambda = \lambda_{x}v$, where~$v$ is the expansion velocity. The advantage of the second measure is that it can be applied in situations when the spatial distribution of the genotypes is available for only a single time point. We were able to compute both fitness measures analytically. The complete details of this calculation are given in the Supplemental Material, but our approach is briefly summarized below.

When a mutant first appears, its abundance is too small to immediately influence the course of the range expansion; therefore, we can study the dynamics of the mutant fraction in the reference frame comoving with the expansion, effectively reducing two coupled equations in Eq.~(\ref{eq:model_main}) to a single equation. This remaining equation has the form of a Fokker-Planck equation with a source term, and its largest eigenvalue determines whether the total fraction of the mutant will increase or decrease with time. We were able to obtain this largest eigenvalue and the corresponding eigenfunction exactly in terms of only elementary functions. For small differences in the dispersal abilities~$|D_1-D_2|\ll D_2$, our result takes a particularly simple form,

\begin{equation}
	\lambda_x = \frac{D_1-D_2}{6 D_{2}}\sqrt{\frac{r}{2D_2}}\left(1-4\frac{c^{*}}{K}\right),
	\label{eq:exact_short}
\end{equation}

\noindent which is valid for~$c^*>-K/2$; see Supplemental Material for~$c^*<-K/2$. Thus,~$\lambda_x$ is a linear function of the Allee threshold~$c^*$, which changes sign at~$c^*=K/4$. For low Allee thresholds, natural selection favors mutants with higher dispersal, but, when growth is highly cooperative, the direction of selection is reversed and slower dispersers are favored. Numerical simulations of Eq.~(\ref{eq:model_main}) are in excellent agreement with our exact solution~(Fig.~\ref{fig:switch}b). In the Supplemental Material, we explain that the direction of natural selection remains the same as the mutant takes over the population and further discuss the effects of mutations and demographic fluctuations by connecting the largest eigenvalue to the fixation probability of the mutant~\cite{supplement}.

\begin{figure}[h!]
\includegraphics[width=0.7\columnwidth]{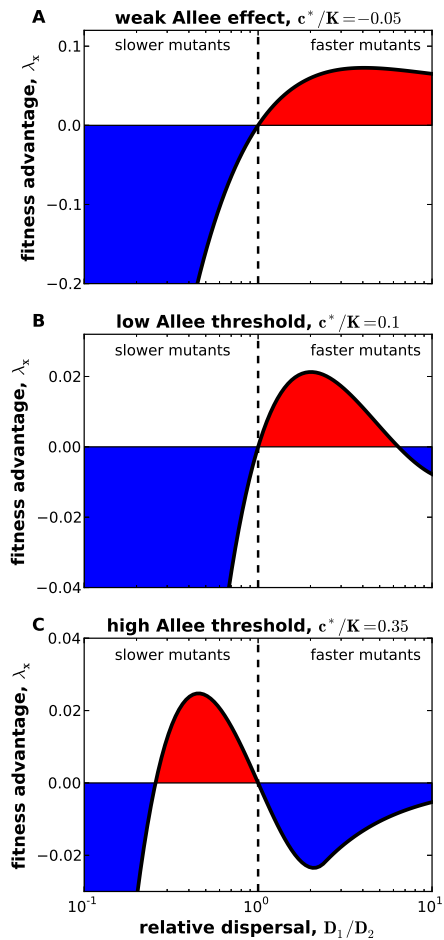}
\caption{Allee effect determines how fitness depends on dispersal. (A) Faster dispersers are unconditionally favored when the Allee effect is weak. Note that the fitness advantage reaches a maximum at a finite~$D_1/D_2$. (B) When the Allee effect is strong, but the Allee threshold is low, selection still favors faster dispersal. Very fast mutants however are at a disadvantage. (C) For high Allee threshold, only slower-dispersing mutants can succeed, but mutants that disperse too slowly are selected against. In all panels, the exact solution is plotted, and colors highlight beneficial~(red) and deleterious~(blue) mutations. The dashed line marks~$D_1=D_2$, where both types have the same fitness. Near this point, the fitness advantage of type one in the background of type two equals the fitness disadvantage of type two in the background of type one, but this symmetry breaks down when the dispersal rates of the types are very different; see Supplemental Material. Nevertheless, the exchange of~$D_1$ and~$D_2$ always converts a beneficial mutant to a deleterious one. In consequence, a mutation that is beneficial when rare will remain beneficial when it approaches fixation indicating that the direction of natural selection is the same for small and large~$f$.}
\label{fig:dependence}
\end{figure}

Our finding that lower dispersal is advantageous seems counterintuitive. Indeed, a mutant unable to disperse cannot possibly take over the expansion front. The resolution of this apparent paradox is that Eq.~(\ref{eq:exact_short}) is only valid for~$D_1\approx D_2$, and the direction of natural selection changes as~$D_1$ approaches zero. The exact expression for the selective advantage for arbitrary~$D_1/D_2$ is given in the Supplemental Material and is plotted in Fig.~\ref{fig:dependence} for different values of the Allee threshold. When the Allee effect is absent or weak, selection unconditionally selects for faster dispersal~(Fig.~\ref{fig:dependence}a), but, as the Allee threshold increases and becomes positive, mutants with very large dispersal rates become less fit than the wild type~(Fig.~\ref{fig:dependence}b). This is expected because mutants that disperse too far ahead of the front cannot reach the critical density necessary to establish a viable population. As a result, there is an optimal improvement in dispersal abilities that is favored by natural selection. In contrast, when the Allee effect is sufficiently strong, only reduced dispersal is advantageous~(Fig.~\ref{fig:dependence}c). Again, there is an optimal reduction in the dispersal rate that results in the highest fitness advantage, and mutants that disperse too slowly are outcompeted by the wild type.

Although natural selection typically eliminates the mutants that either increase or decrease the dispersal rate by a large amount, sequential fixation of mutations could lead to a substantial change in the expansion velocity. Indeed, our results show that the fitness advantage of the mutant depends on the relative rather than absolute change in the dispersal ability. Thus, if the Allee effect is strong enough to favor slower mutants, then mutants that reduce the dispersal rate even further will become advantageous once the takeover by the original mutant is complete. We then expect that the repeated cycle of dispersal reduction will eventually bring the invasion to a standstill. The opposite behavior is expected when the Allee threshold is low. 

\begin{figure}[h!]
\includegraphics[width=\columnwidth]{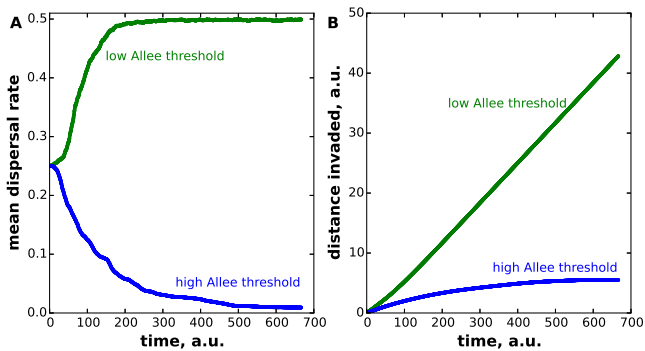}
\caption{When the rate of dispersal is allowed to evolve, simulations show that invasions can both accelerate and decelerate depending on the strength of an Allee effect. (A) The mean dispersal rate increases to its maximally allowed value when the Allee threshold is low~(green), but the dispersal rate decreases to zero when the Allee threshold is high~(blue). (B) For the same simulations as in (A), we plot the extent of spatial spread by the populations. Invasions with a low Allee threshold~($c^*/K=0.2$) accelerate, while invasions with a high Allee threshold~($c^*/K=0.35$) come to a standstill.}
\label{fig:arrest}
\end{figure}

% expansion arrest in simulations
To test these predictions, we performed computer simulations that relax many of the assumptions underlying Eq.~(\ref{eq:model_main}) as described in the Supplemental Material. In particular, we incorporated the stochastic fluctuations due to genetic drift and allowed multiple mutations modifying the dispersal rate to arise and compete at the same time. Shown in Fig.~\ref{fig:arrest}, simulations display a steady decline in the dispersal ability and expansion arrest for strongly cooperative growth. Consistent with previous studies~\cite{thomas:acceleration, shine:spatial_sorting, korolev:wave_splitting, benichou:acceleration}, dispersal rates increase and the rate of invasion accelerates when the Allee threshold is low.

%**************************************************************************
% Discussion
%**************************************************************************

Natural selection on dispersal has been extensively studied, and many factors that favor faster or slower dispersal have been identified~\cite{ronce:dispersal_review,dockery:slow_dispersal,hutson:slow_dispersal}. Fast dispersers can avoid inbreeding depression, escape competition, or find a suitable habitat. At the same time, dispersal diverts resources from reproduction and survival, increases predation, and can place organisms in inhospitable environments. In the context of range expansions, however, high dispersal seems unambiguously beneficial because early colonizers get a disproportionate advantage. Yet, we showed that spatial expansions can select for mutants with lower dispersal rates. Continuous reduction of dispersal rates then slows down and eventually stops further invasion. Invasion arrest requires strong cooperative growth and is in stark contrast to the dynamics in non-cooperative populations where spatial expansions select for higher dispersal.

We expect that our results are robust to the specific assumptions made in this study such as the diffusion-like dispersal and the specific form of the growth function because, at its core, our analysis relies on very general arguments~(Supplemental Material). Indeed, we argue that faster mutants get ahead at low or negative Allee thresholds because they can successfully grow at the front and effectively establish secondary invasions; in contrast, these dynamics do not occur at high Allee thresholds because faster dispersers arrive at low-density regions that cannot sustain growth.

At a very high level, our result can be understood as the emergence of cheating in a cooperatively growing population. Cheating is a behavior that benefits the individuals, but is detrimental to the population as a whole~\cite{west:social}. One well-studied example is consuming, but not contributing, to a common resource~(public good), a behavior typical of both humans~\cite{hardin:tragedy, axelrod:cooperation, gardner:cooperation_history} and microbes~\cite{west:social, nadell:sociobiology, menon:diffusion}. In the context of population spreading, high dispersal can be viewed as an effective public good because it creates high densities in the outer edge of the expansion front, thereby increasing the survival of new immigrants to that region. Although high population densities benefit both slow and fast dispersers equally, the latter pay a much higher cost for producing this public good. Indeed, faster dispersers are more likely to suffer higher death rates at the low-density invasion front, where they arrive more frequently. As a result, ``cheating'' by the slow dispersers is the reason for their selective advantage.

In addition to the classical emergence of cheating, expansion arrest is an example of evolution driving a population to a less adapted state. Our ability to exploit or trigger such counterproductive evolution may be important in managing invasive species and agricultural pests, or even cancer tumors. Concretely, our results open up new opportunities to control biological invasions. Instead of trying to kill the invader, a better strategy could be to elevate the minimal density required for growth~(the Allee threshold) to a level necessary for evolution to select for invasion arrest. Such strategies could have important advantages over the traditional approaches. Increasing the Allee threshold in cancer tumors could overcome the emergence of drug resistance because the bulk of the tumor is at a high density and is not affected by the treatment. Similarly, resistance should emerge much more slowly in agricultural pests.

Although the manipulation of Allee thresholds is a relatively unexplored and potentially difficult endeavor, some management programs have been successful at increasing the Allee effects in the European gypsy moth, one of the most expensive pests in the United States~\cite{tobin:moth_review, tobin:slow_spread, thorpe:mating_disruption, johnson:allee_gypsy}. These moths suffer from a strong Allee effect because they struggle to find mates at low population densities~\cite{johnson:allee_gypsy}. Recent management programs exacerbated this Allee effect by spreading artificial pheromones that disorient male moths and prevent them from finding female mates, thereby effectively eradicating low-density populations~\cite{tobin:moth_review, tobin:slow_spread, thorpe:mating_disruption}. European gypsy moths and other pests with similarly strong Allee effects could be close to the critical Allee threshold necessary for the invasion arrest. In such populations, further increase in of the Allee effect could be more feasible and effective than reducing the carrying capacity. 

Beyond ecology, our results could also find applications in other areas of science such as chemical kinetics, where reaction-diffusion equations are often used. Quite broadly, we find that a variation in the motility of agents can have completely opposite effects on their dynamics depending on the reaction kinetics at the expansion front.

\begin{acknowledgments}
This work was supported by the startup fund from Boston University to KK. Simulations were carried out on Shared Computing Cluster at BU.
\end{acknowledgments}

\clearpage 

\onecolumngrid
% removes paragraph indentation
\setlength{\parindent}{0in}
% set spacing between paragraphs to 1 line
\parskip = \baselineskip

\renewcommand{\thefigure}{S\arabic{figure}}
\setcounter{figure}{0}
\renewcommand{\theequation}{S\arabic{equation}}
\setcounter{equation}{0}

\textbf{\Large Supplemental Material for ``Evolution arrests invasions of cooperative populations''.}\\

%*************************************************************************
% Model formulation
%*************************************************************************
\textbf{\large Formulation of the mathematical model}

We consider the competition of two types~(mutants, strains, or species) with different dispersal abilities as they invade new territory. As defined in the main text, the model is expressed as

\begin{equation}
	\begin{aligned}
		& \frac{\partial c_{1} }{\partial t} = D_{1}\frac{\partial^2 c_{1}}{\partial x^2} + c_{1}g(c),\\
		& \frac{\partial c_{2} }{\partial t} = D_{2}\frac{\partial^2 c_{2}}{\partial x^2} + c_{2}g(c).
	\end{aligned}
	\label{eq:model}
\end{equation}

\noindent With the following expression for~$g(c)$:

\begin{equation}
	g(c)=r(K-c)(c-c^{*})/K^{2}. 
	\label{eq:growth}
\end{equation}

In the following, we assume that type two is the wild type and is in the majority while type one is a recent mutant with a different dispersal rate. Our goal is to compute the fitness advantage of type one and determine whether selection favors slower or faster dispersal.

%**************************************************************************
% Fitness advantage as an eigenvalue problem
%**************************************************************************
\textbf{\large Fitness advantage is a solution of an eigenvalue problem}

The fate of mutations is determined when their density is quite small~\cite{gillespie:book}. Therefore, we assume that the population density of the first type is much lower than that of the second type, whose dispersal and growth determine the dynamics of the expansion. We then compute whether the fraction of the first type increases in time indicating that it will eventually take over or decreases with time indicating that it is less fit and will be eventually eliminated by natural selection. To achieve this goal, it is convenient to represent population dynamics in terms of the total population density~$c$ and the fraction of the first type~$f$ defined below as

\begin{equation}
	\begin{aligned}
		& c(t,x) = c_{1}(t,x) + c_{2}(t,x),\\
		& f(t,x) = \frac{c_{1}(t,x)}{c_{1}(t,x) + c_{2}(t,x)}.
	\end{aligned}
	\label{eq:fc}
\end{equation}

This change of variables results in the following set of reaction-diffusion equations:

\begin{equation}
	\begin{aligned}
		& \frac{\partial c}{\partial t} = D_{2}\frac{\partial^2 c}{\partial x^2} + cg(c) + (D_1-D_2)\frac{\partial^2 (fc)}{\partial x^2} ,\\
		& \frac{\partial f}{\partial t} = D_{1}\frac{\partial^2 f}{\partial x^2} + \frac{2D_1}{c}\frac{\partial c}{\partial x}\frac{\partial f}{\partial x} + f \frac{D_1-D_2}{c}\frac{\partial^2 c}{\partial x^2} - f\frac{D_1-D_2}{c}\frac{\partial^2 (fc)}{\partial x^2}.
	\end{aligned}
	\label{eq:f_dynamics_full}
\end{equation}

\noindent Since we assume that~$f\ll 1$, we neglect higher order terms in~$f$, i.e. the terms linear in~$f$ in the equation for the population density and terms quadratic in~$f$ in the equation for the fraction of the first type. The result reads

\begin{equation}
	\begin{aligned}
		& \frac{\partial c}{\partial t} = D_{2}\frac{\partial^2 c}{\partial x^2} + cg(c),\\
		& \frac{\partial f}{\partial t} = D_{1}\frac{\partial^2 f}{\partial x^2} + \frac{2D_1}{c}\frac{\partial c}{\partial x}\frac{\partial f}{\partial x} + f \frac{\Delta D}{c}\frac{\partial^2 c}{\partial x^2},
	\end{aligned}
	\label{eq:f_dynamics}
\end{equation}

\noindent where we introduced~$\Delta D = D_1 - D_2$. The equation for~$c$ is now independent from~$f$ and its solution is known exactly for the quadratic form of~$g(c)$ with~$c^{*}>-K/2$~\cite{aronson:allee_wave, fife:allee_wave}:

\begin{equation}
	\begin{aligned}
		&v=\sqrt{\frac{D_2r}{2}}\left(1-2\frac{c^{*}}{K}\right),\\
		&c(\zeta)=\frac{K}{1+e^{\sqrt{\frac{r}{2D_2}}\zeta}},
	\end{aligned}
	\label{eq:profile}
\end{equation}
\noindent where~$\zeta=x-v_{2}t$ in the spatial coordinate in the reference frame comoving with the expansion front. Note that for~$c^*>K/2$ the velocity is negative and the whole population is driven to extinction. For~$c<-K/2$, the cooperative growth has a negligible effect on the expansion dynamics and~$v=2\sqrt{Dr|c^{*}/K|}$ and~$c(\zeta) \sim \exp(-\zeta\sqrt{r|c^{*}|/K})$ for large~$\zeta$, which are obtained by the linearization of the growth dynamics in~$c$ as for the classic Fisher-Kolmogorov equation~\cite{fisher:wave, kolmogorov:wave, saarloos:review}.

As a result of this simplification, we treat the equation for~$f$ as a single partial differential equation with non-constant, but known, coefficients. To solve this equation, we change to the reference frame comoving with the population expansion, which is defined by the following change of variables

\begin{equation}
	\begin{aligned}
		& \zeta = x -v_{2}t,\\
		& \tau = t.
	\end{aligned}
	\label{eq:comoving}
\end{equation}

\noindent Note that the velocity of the comoving frame is given by equation~(\ref{eq:profile}) with the parameters of the second type, which is in the majority. To indicate that, we denote this velocity as~$v_2$. We also denote the time as~$\tau$ to emphasize the change of variables. In the comoving reference frame, the equation for~$f$ reads

\begin{equation}
	\frac{\partial f}{\partial \tau} = D_{1}\frac{\partial^2 f}{\partial \zeta^2} + \left(v_2 + \frac{2D_1}{c}\frac{\partial c}{\partial \zeta}\right)\frac{\partial f}{\partial \zeta} + f \frac{\Delta D}{c}\frac{\partial^2 c}{\partial \zeta^2}.
	\label{eq:f_comoving}
\end{equation}

Since equation~(\ref{eq:f_comoving}) is a linear partial differential equation of the form

\begin{equation}
	\frac{\partial f}{\partial \tau} = \mathbf{L} f,
	\label{eq:operator_L}
\end{equation}

\noindent the long term dynamics and the eventual fate of the newly introduced type is determined by~$\lambda$, the largest eigenvalue of~$\mathbf{L}$. When~$\Delta D=0$, the largest eigenvalue equals zero and corresponds to an eigenvector that does not depend on~$\zeta$. Indeed, the system must relax to a spatially homogeneous~$f(\tau, \zeta)$ when the two type are identical. When~$\Delta D \ne 0$, one of the types outcompetes the other at rate~$\lambda$. Positive~$\lambda$ correspond to the takeover by the first type while negative~$\lambda$ indicates that the first type is less fit and will be eliminated. Thus, our goal is to solve the following eigenvalue problem for the largest possible~$\lambda$

\begin{equation}
	\lambda f= D_{1}f'' + \left(v_2 + 2D_1\frac{c'}{c}\right)f' + f \Delta D\frac{c''}{c},
	\label{eq:f_eigen_value}
\end{equation}

\noindent where the derivatives with respect to~$\zeta$ are denoted with primes.

%**************************************************************************
% Exact solution of the eigenvalue problem 
%**************************************************************************
\textbf{\large Exact solution of the eigenvalue problem}

We now proceed by solving equation~(\ref{eq:f_comoving}) exactly. To that end, it is convenient to recast equation~(\ref{eq:f_eigen_value}) in a Hermitian form by eliminating the gradient term on the right hand side. This is accomplished by the following change of variables

\begin{equation}
	\begin{aligned}
		& f(\zeta) = \psi(\zeta) e^{u(\zeta)}, \;\text{where}\\
		& u(\zeta) = -\frac{v_{2}\zeta}{2D_{1}} - \ln(c(\zeta)).
	\end{aligned}
	\label{eq:u}
\end{equation}

\noindent The resulting eigenvalue problem then reads

\begin{equation}
	\lambda \psi = D_1 \psi''- \psi\left(D_2\frac{c''}{c} + v_2 \frac{c'}{c} + \frac{v_{2}^{2}}{4D_1} \right), 
	\label{eq:eigen_value_hermitian}
\end{equation}

The potential term can be further simplified by using the density equation in equations~(\ref{eq:f_dynamics}) written in a comoving reference frame~($-v_{2}c'=D_{2}c''+g(c)c$), with the following result:

\begin{equation}
	\lambda \psi = D_1 \psi'' + \psi\left(g(c(\zeta)) - \frac{v_{2}^{2}}{4D_1} \right). 
	\label{eq:eigen_value_hermitian_simplified}
\end{equation}

\noindent Since the solution techniques differ depending on whether~$c^*<-K$, $c^*\in(-K,-K/2)$, or $c^*>-K/2$, we analyze these three cases separately.

For~$c^{*}<-K$, there is no Allee effect, and~$g(c)$ is a decaying function of the population density. Therefore, the potential term in equation~(\ref{eq:eigen_value_hermitian_simplified}) reaches its maximum when~$c\to0$ and~$\zeta\to+\infty$. Moreover, since the potential term is constant for large~$\zeta$, we conclude that the eigenfunction corresponding to the largest eigenvalue is concentrated at large~$\zeta$, and the eigenvalue itself is given by the maximal value of the potential:

\begin{equation}
	\lambda = r\left(1-\frac{1}{p}\right)\left|\frac{c^*}{K}\right|,
	\label{eq:lambda_exact_fisher}
\end{equation}

where we defined~$D_1/D_2=p$ for convenience. Thus, faster dispersing types are unconditionally favored when cooperation plays no role in population dynamics. The maximal fitness advantage a mutant can obtain by increasing its dispersal rate is~$r|c^{*}|/K$, while the cost of reduced motility can be infinitely large. This asymmetry is due to the fact that we observe changes in the frequency of type one relative to type two, who is in the majority. As a result, when a new mutant is not moving it is immediately lost from the front. In contrast, when the mutant is spreading rapidly, it still takes an appreciable time for the density of this mutant to reach high values at the expanding front of the wild type, even though the mutant almost immediately outcompetes the wild type at the far edge of the expansion~($\zeta\to+\infty$).

From our analysis of~$c^*>-K/2$, it will be clear that faster dispersers are also unconditionally favored for~$c^*\in(-K,-K/2)$ and equation~(\ref{eq:lambda_exact_fisher}) holds for~$p\ge p_{+\infty}$, where~$p_{+\infty}$ decreases from~$1$ at~$c^{*}=-K/2$ to~$0$ at~$c^*=-K$. For~$p < p_{+\infty}$, the slower mutant will be lost less rapidly than predicted by equation~(\ref{eq:lambda_exact_fisher})  because it can take advantage of the higher per capita growth rates in the interior of expansion front~($g(c)$ is not monotonically decreasing in this regime). Nevertheless, we find that equations (\ref{eq:lambda_exact_fisher}) and (\ref{eq:lambda_exact}) provide a good approximation to the decay rates for~$c^*\in(-K,-K/2)$ and~$p < p_{+\infty}$.

We now turn to~$c^*>-K/2$. Since, in this regime, the shape of density profile is known exactly, it is convenient to perform a change of variable~$\rho = c(\zeta)/K$ in equation~(\ref{eq:f_eigen_value}) and treat~$\rho$ as an independent variable, thus eliminating~$\zeta$. The resulting eigenvalue problem reads

\begin{equation}
\frac{rp}{2}\rho^2(1-\rho)^2 f'' + \frac{rp}{2} \rho(1-\rho)\left[3-4\rho-\frac{1-2\rho^*}{p}\right] f' + \frac{r}{2}(p-1)(1-\rho)(1-2\rho)f = \lambda f,
	\label{eq:eigen_rho}
\end{equation}

\noindent where the primes now denote the derivatives with respect to~$\rho$, and~$\rho^*=c^*/K$. We further simplify equation~(\ref{eq:eigen_rho}) by letting~$f(\rho) = h(\rho)e^{\nu(\rho)}$ and choosing~$\nu(\rho)$ to set the term with~$h'$ to zero. This results in

\begin{equation}
	\nu = -\frac{3}{2}\ln\rho - \frac{1}{2}\ln(1-\rho) + \frac{(1-2\rho^*)}{2p}\ln\frac{\rho}{1-\rho}, 
	\label{eq:nu}
\end{equation}

\begin{equation}
	p\rho^2(1-\rho)^2 h'' + \left[-2\rho^2 + 2(1+\rho^*)\rho + \frac{p}{4} - 2\rho^* - \frac{(1-2\rho^*)^2}{4p} \right]h = \frac{2\lambda}{r}h
	\label{eq:eigen_rho_pseudohermitian}
\end{equation}

\noindent To convert this equation into a Hermitian form, we divide both sides by~$\rho(1-\rho)$ and define~$\varphi=h/[\rho(1-\rho)]$:

\begin{equation}
	p\rho(1-\rho)\frac{d^2}{d\rho^2}[\rho(1-\rho)\varphi] + \left[-2\rho^2 + 2(1+\rho^*)\rho + \frac{p}{4} - 2\rho^* - \frac{(1-2\rho^*)^2}{4p} \right]\varphi = \frac{2\lambda}{r}\varphi.
	\label{eq:eigen_rho_hermitian}
\end{equation}

\noindent The derivative term is now clearly Hermitian because it acts symmetrically to the left and to the right, which was not the case in equation~(\ref{eq:eigen_rho_pseudohermitian}). 

It is now apparent that we should look for~$\varphi=\rho^{\alpha-1}(1-\rho)^{\beta-1}$ because

\begin{equation}
	[\rho^{\alpha}(1-\rho)^{\beta}]''= \rho^{\alpha-2}(1-\rho)^{\beta-2}[(\alpha+\beta)(\alpha+\beta-1)\rho^2 - 2\alpha(\alpha+\beta-1)\rho + \alpha(\alpha-1)].
\label{eq:derivative_ansatz}
\end{equation}

\noindent Indeed, for such a choice of~$\varphi$, the first term becomes a product of~$\varphi$ and a quadratic polynomial of~$\rho$, which is exactly the form of the other terms in the equation. Moreover, this ansatz yields a unique solution because there are three unknowns~$\alpha$,~$\beta$, and~$\lambda$ and three coefficients of the quadratic polynomial to match. We then find that the values of~$\alpha$ and~$\beta$ that satisfy equation~(\ref{eq:eigen_rho_pseudohermitian}) are given by

\begin{equation}
	\begin{aligned}
		& \alpha = \frac{1+\rho^*}{4}\left(1+\sqrt{1+\frac{8}{p}}\right),\\
		& \beta = \frac{1-\rho^*}{4}\left(1+\sqrt{1+\frac{8}{p}}\right),\\
	\end{aligned}
	\label{eq:ab_solution}
\end{equation}

\noindent for

\begin{equation}
\lambda = \frac{r}{2}\left[p\left(\alpha-\frac{1}{2}\right)^2 - 2\rho^* - \frac{(1-2\rho^*)^2}{4p} \right].
	\label{eq:lambda_exact}
\end{equation}
 
\noindent Note that the eigenvalue that we found is the largest because the corresponding eigenfunction has no zeros for~$\rho\in(0,1)$~\cite{titchmarsh:second_order_ode}.

To ensure that equation~(\ref{eq:lambda_exact}) indeed describes the fitness advantage of a mutant, we also need to check that the corresponding eigenfunction has a finite~$\mathbf{L}^2$ norm; otherwise, it cannot serve as a basis vector in the Hilbert space. The integrability of~$\varphi^2$ requires that~$\alpha>1/2$ and~$\beta>1/2$. We will discuss the consequences of these requirements for positive and negative~$\rho^*$ separately.

For~$\rho^*\le0$, the condition on~$\beta$ is always satisfied, while~$\alpha>1/2$ only when

\begin{equation}
	p<p_{+\infty}=\frac{2(1+\rho^*)^2}{-\rho^*}.
\label{eq:p_alpha}
\end{equation}

\noindent As~$p$ approaches~$p_{+\infty}$ from below, the eigenfunction becomes localized at~$\rho=0$, which corresponds to~$\zeta\to+\infty$. Above~$p_{+\infty}$, equation~(\ref{eq:lambda_exact}) becomes invalid, and the eigenvalue is given by the limit of the potential term in equation~(\ref{eq:eigen_value_hermitian_simplified}) as~$\zeta\to+\infty$. The result reads

\begin{equation}
	\lambda_{+\infty}=\lim_{\zeta\to+\infty}\left(g(c(\zeta)) - \frac{v_{2}^{2}}{4D_1}\right) = - r\rho^* - \frac{v_{2}^{2}}{4D_1} = - r\rho^* - \frac{r(1-2\rho^*)^2}{8p}>0.
	\label{eq:lambda_alpha}
\end{equation}

Since~$p_{+\infty}$ decreases from~$+\infty$ to~$1$ as~$\rho^*$ changes from~$0$ to~$-1/2$, we expect that, for~$\rho^*<-1/2$, the fitness advantage of the mutant is given by~$\lambda_{+\infty}$ for~$p>1$ as well as for~$p$ above some critical value, which we also label~$p_{+\infty}$ to extend our definition of this quantity to~$\rho^*<-1/2$. Note that, although we have not shown that this extension obeys equation~(\ref{eq:p_beta}), this equation might still provide a good approximation for~$p_{+\infty}$  given that it predicts that~$p_{+\infty}$ approaches~$0$ as~$\rho^*$ approaches~$-1$. 

For~$\rho^*\ge0$, the condition on~$\alpha$ is always satisfied, while~$\beta>1/2$ only when

\begin{equation}
	p<p_{-\infty}=\frac{2(1-\rho^*)^2}{\rho^*}.
\label{eq:p_beta}
\end{equation}

\noindent As~$p$ approaches~$p_{-\infty}$ from below, the eigenfunction becomes localized at~$\rho=1$, which corresponds to~$\zeta\to-\infty$. Above~$p_{-\infty}$, equation~(\ref{eq:lambda_exact}) becomes invalid, and the eigenvalue is given by the limit of the potential term in equation~(\ref{eq:eigen_value_hermitian_simplified}) as~$\zeta\to-\infty$. The result reads

\begin{equation}
	\lambda_{-\infty}=\lim_{\zeta\to-\infty}\left(g(c(\zeta)) - \frac{v_{2}^{2}}{4D_1}\right)=- \frac{v_{2}^{2}}{4D_1} = - \frac{r(1-2\rho^*)^2}{8p}<0.
	\label{eq:lambda_beta}
\end{equation}

Collectively, equations~(\ref{eq:lambda_exact}), (\ref{eq:lambda_alpha}), (\ref{eq:lambda_beta}) completely specify the exact solution for the fitness advantage of the mutant for~$\rho^*\in(-1/2,1/2)$. Unless specified otherwise, we will denote this solution as simply~$\lambda$ regardless of whether it is specified by~(\ref{eq:lambda_exact}) or equations (\ref{eq:lambda_alpha}) and (\ref{eq:lambda_beta}). The behavior of~$\lambda$ as a function of~$\rho^*=c^*/K$ and~$p=D_1/D_2$ is discussed next.

For~$c^*/K\in(-1/2,0)$, faster dispersal is always advantageous similar to the results for~$c^*<-K/2$. This result is expected because populations with negative~$c^*$ do not require a critical density for growth, and, thus, an organism dispersing very far ahead of the invasion front can still establish a viable population. Although positive, the selective advantage of faster dispersers does not increase monotonically with~$D_1/D_2$ as it does for populations without an Allee effect. Instead,~$\lambda$ has a maximum at a finite value of~$p$ as shown in Fig.~2A in the main text.

For~$c^*/K\in(0,1/4)$, we expect that very fast dispersers have a negative selective advantage because they cannot establish a viable population due to the strong Allee effect. Consistent with this expectation, equation~(\ref{eq:lambda_exact}) predicts that~$\lambda$ is negative for~$D_1<D_2$, zero for~$D_1=D_2$, and positive for~$D_1/D_2\in(1,p_{0})$, where~$p_{0}$ is another root of~$\lambda(p)$ and is given by

\begin{equation}
	p_{0} = \frac{(1-2\rho^*)^2}{\rho^*}.
	\label{eq:p_0}
\end{equation}

\noindent As~$p$ is increased beyond~$p_0$, the fitness advantage~$\lambda$ becomes negative. In summary, faster dispersers do have a higher fitness unless they disperse too far ahead and suffer from the strong Allee effect.

Although equation~(\ref{eq:lambda_exact}) predicts another zero of~$\lambda(p)$ at

\begin{equation}
	p_1 = \frac{2(2\rho^*-1)^2}{3-5\rho^*-\sqrt{9-34\rho^*+41(\rho^{*})^2-16(\rho^*)^3}},
	\label{eq:p_1}
\end{equation}

\noindent and positive~$\lambda$ for~$D_1>p_1D_2$, we find that~$p_1>p_{-\infty}$ and, thus, the eigenvalue is given by~$\lambda_{-\infty}$, which is negative. Our simulations confirm that~$\lambda$ is negative for all~$D_1>p_0D_2$. 

The remaining region,~$c^*/K\in(1/4,1/2)$, favors slower dispersers with~$D_1/D_2\in(p_0,1)$ while mutants that disperse either too slowly or faster than the wild type have a negative selective advantage. Note that as~$\rho^*$ is increased above~$1/4$,~$p_0$ becomes less than $1$ and the region of~$D_1/D_2$ that is favored by natural selection shifts from just above~$1$ to just below~$1$. Similar to the situation we just discussed,~$p_1>p_{-\infty}$, and, thus,~$\lambda$ remain negative for all~$D_1>D_2$.

\begin{figure}
\includegraphics[width=0.5\columnwidth]{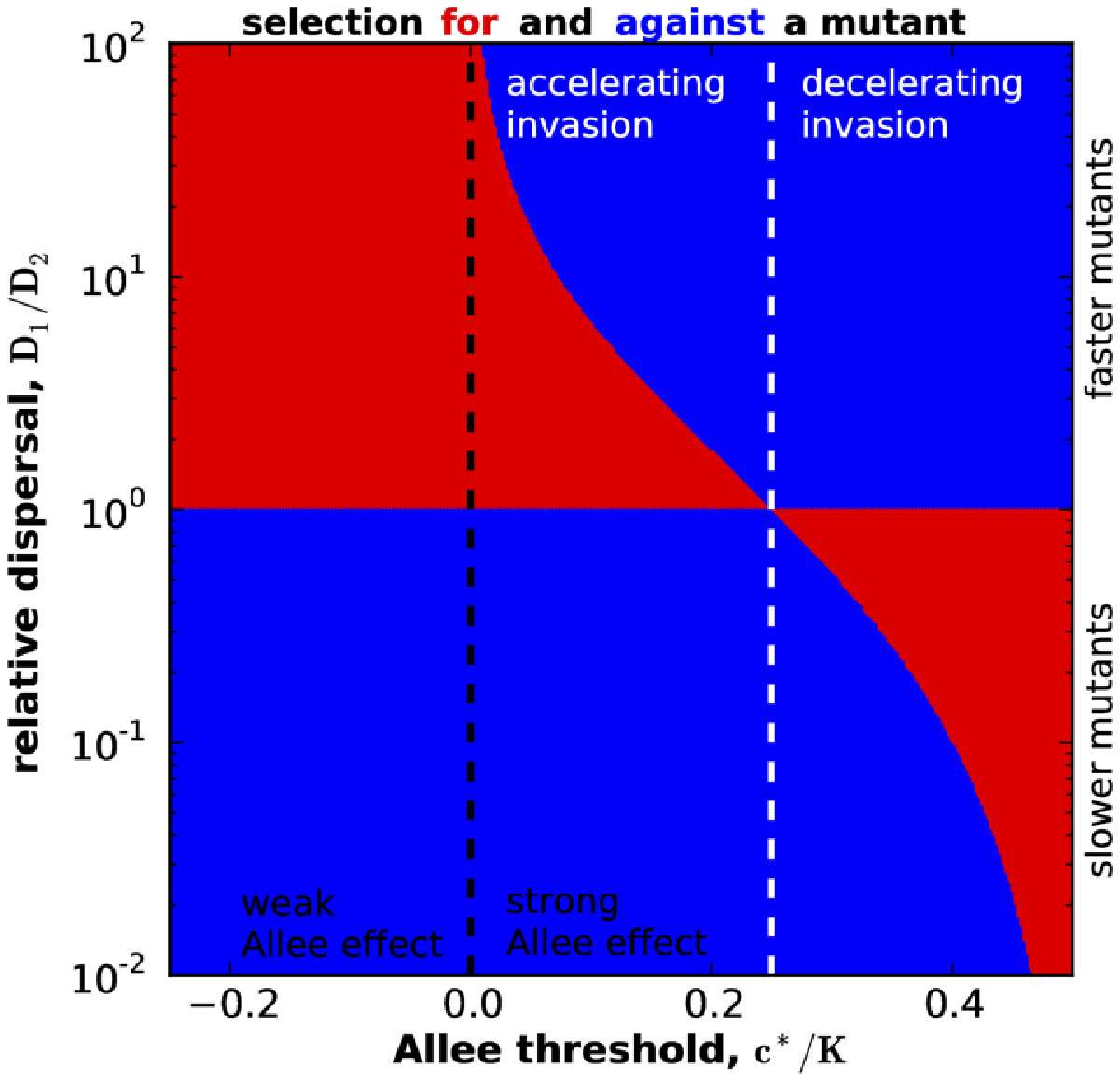}
\caption{The fate of a mutant depends both on its dispersal abilities and the Allee effect in the population growth. Advantageous mutants are in red and deleterious are in blue. Note that the exchange of the types corresponds to an inversion in the point~$(c^*/K=1/4,\;p=1)$ only for~$D_1\approx D_2$. This phase diagram is drawn based on the exact solution described in this Supplemental Material. }
\label{fig:phase_diagram}
\end{figure}

These results are summarized graphically in Fig.~\ref{fig:phase_diagram}, which shows the regions in the space of~$\rho^*$ and~$p$ favoring faster or slower dispersal. 

Finally, we analyze how the density profile of the mutant depends on~$\zeta$. The shape of this density profile can be obtained at a single time point and, therefore, could be of great utility in practice because it contains information about the fitness advantage of a mutant and does not require time series data. To compute~$f(\zeta)$ we need to combine the eigenfunction that we found above as well as the transformations from~$f$ to~$\varphi$; the result reads: 

\begin{equation}
	f \propto \rho^{\alpha-\frac{3}{2}+\frac{1-2\rho^*}{2p}}(1-\rho)^{\beta-\frac{1}{2}-\frac{1-2\rho^*}{2p}}.
	\label{eq:f_shape}
\end{equation}

For negative~$c^*$, $f(\zeta)$ is monotonically increasing from~$0$ (at~$\zeta=-\infty$) to~$+\infty$ (at~$\zeta=+\infty$) for~$D_1>D_2$ and monotonically decreasing from~$+\infty$ (at~$\zeta=-\infty$) to~$0$ (at~$\zeta=+\infty$) for~$D_1<D_2$.

For~$\rho^*\in(0,1/4)$, the behavior is the same as above for~$D_1<D_2$ and~$D_1\in(1,p_0D_2)$. For~$D_1>p_0D_2$, the profile of~$f(\zeta)$ has a minimum in the region of the front, i.e. around~$\zeta=0$, and diverges to~$+\infty$ as~$\zeta\to\pm\infty$.

For~$\rho^*\in(1/4,1/2)$, $f(\zeta)$ monotonically decreases from~$+\infty$ (at~$\zeta=-\infty$) to~$0$ (at~$\zeta=+\infty$) for~$D_1<p_0D_2$, then, for~$D_1/D_2\in(p_0,1)$ it becomes peaked around the front, i.e. close to~$\zeta=0$, and decays to~$0$ as~$\zeta\to\pm\infty$, while, for~$D_1>D_2$,~$f(\zeta)$ has a minimum around~$\zeta=0$ and diverges to~$+\infty$ as~$\zeta\to\pm\infty$.

%**************************************************************************
% Extras
%**************************************************************************
\textbf{\large Supplemental Discussion}

\textit{Effects of different growth rates}\\
Mutations that change the rate of dispersal can also affect the growth rate. Although the dispersal-survival and dispersal-fecundity trade-offs have been most heavily documented, other possibilities exist including potentially a reduction in feeding ability due to limited dispersal~\cite{tobin:moth_review}. Here, we outline how the differences in growth rates can be included in our theory and show that, under certain simplifying assumptions, the net rate of increase of a mutant is simply given by the sum of the difference in the growth rates and~$\lambda$ due to the differences in the dispersal abilities, which we obtained above.

When types grow at different rates~$g_{1}(c_1,c_2)$ and~$g_{2}(c_1,c_2)$, Eq.~(\ref{eq:model}) needs to be modified as

\begin{equation}
	\begin{aligned}
		& \frac{\partial c_{1} }{\partial t} = D_{1}\frac{\partial^2 c_{1}}{\partial x^2} + c_{1}g_{1}(c_1,c_2),\\
		& \frac{\partial c_{2} }{\partial t} = D_{2}\frac{\partial^2 c_{2}}{\partial x^2} + c_{2}g_{2}(c_1,c_2).
	\end{aligned}
	\label{eq:model_dgr}
\end{equation}

\noindent Assuming~$f\ll1$ and repeating the steps leading to Eq.~(\ref{eq:f_comoving}), we obtain

\begin{equation}
	\frac{\partial f}{\partial \tau} = D_{1}\frac{\partial^2 f}{\partial \zeta^2} + \left(v_2 + \frac{2D_1}{c}\frac{\partial c}{\partial \zeta}\right)\frac{\partial f}{\partial \zeta} + f \frac{\Delta D}{c}\frac{\partial^2 c}{\partial \zeta^2} + f(g_{1}(c)-g_{2}(c)),
	\label{eq:f_comoving_dgr}
\end{equation}

\noindent which leads to a new eigenvalue problem with an additional potential term due to~$g_{1}(c)-g_{2}(c)\ne0$. Note that the new term depends only on the total population density because~$c_1\ll c_2$.

The solution to this problem depends on the functional form of the difference in the growth rates. It is important to emphasize that, generically, the difference in the growth rate is not a number, but rather a function of the population density. This fact substantially complicates the analysis because greater dispersal can come at the cost of lower growth rates at low population densities or at high population densities. Even more complicated, a mutant can have a reduced growth rate at low densities, but an increased growth rate at high densities. Therefore, we will limit our discussion to two general observations.

First, if~$g_{1}(c)-g_{2}(c)=s$, where~$s$ is a constant fitness advantage, then the additional potential term is just a constant, and the resulting eigenvalue is given by the sum of~$s$ and~$\lambda$ obtained previously. Thus, for density-independent changes in the growth rate, the different components of fitness simply add.

Second, when the difference in the growth rates is not constant, its contribution to the eigenvalue will be given by the average of the new term in the potential over the corresponding eigenfunction. Therefore, we can be certain that an increase in the growth rate leads to a higher eigenvalue while a decrease in the growth rate leads to a lower eigenvalue.

In sum, the effect of different growth rates is more intuitive than the effect of different dispersal rates, and, in the simplest situation, one can just add the different fitness components.

\textit{The direction of natural selection at non-vanishing mutant fractions}\\
In our simulations, we found that the sign of~$\lambda$ fully determines the fate of the mutant even when its fraction becomes no longer small to justify the linearization, which we used to obtain the analytical solution. In particular, we never observed the coexistence of two types due to a change in the direction of natural selection as~$f$ grows from~$0$ to~$1$. For mutants with~$\lambda>0$, these findings can be understood by considering population dynamics when the mutant fraction approaches~$1$. When~$f$ is large, the relative fraction of the ancestor is small, and it is the dispersal rate of the mutant that determines the expansion velocity and the shape of the expansion front. Therefore, the roles of the mutant and the ancestor are reversed, and we can obtain the growth rate of the ancestor from our exact solution by exchanging~$D_1$ and~$D_2$. As illustrated in Fig.~3, the swap of the dispersal rates always converts a positive~$\lambda$ to a negative~$\lambda$. In consequence, the ancestor will have a negative growth rate when the beneficial mutant is in the majority, i.e. the direction of natural selection remains the same for both small and large~$f$.

A more intuitive explanation comes from Fig.~\ref{fig:phase_diagram}, which shows that whether faster or slower dispersal is favored depends only on~$c^*/K$. Therefore, only the magnitude, but not the direction of the selection changes as the mutant becomes established. At intermediate mutant frequencies, the increase in~$f$ occurs due to the competition between the new mutant offspring with the dispersal rate~$D_1$ and the resident population~(consisting of both the mutant and the ancestor) with the dispersal rate qualitatively given by~$fD_1+(1-f)D_2$. Since the difference between the dispersal rate of the mutant and effective dispersal rate of the resident population remains of the same sign, the mutant offspring continue to outcompete the ancestor as~$f$ increases.

We also note that mutants with negative~$\lambda$ do not reach high enough population densities to violate our linearization assumption, so  their dynamics are also fully determined by the direction of selection at low mutant frequencies. 

\textit{Stochastic effects due to mutations and genetic drift}\\
The analytical solution that we presented in this Supplemental Material does not account for the effects of mutations and the stochastic effects due to the randomness of births and deaths. Our simulations, however, include both of these processes, and the agreement between the theory and the simulations demonstrates that our conclusions are robust to the vagaries of mutations and genetic drift. Here, we explain why our deterministic theory is sufficient to capture the essence of the evolutionary process.

We first consider the effects of mutations. Mutations that create disadvantageous types lead to a background level of genotypes with lower fitness similar to the mutation-selection balance in well-mixed populations~\cite{gillespie:book}. Repeated advantageous mutations also have a negligible effect on the individual dynamics of these mutants. Indeed, the population densities of these mutants do not affect each other's growth because the mutants are rare initially and compete with the resident type rather than with each other. This decoupling of mutant evolution will cease once the densities of the mutants become large, and we expect that the fitter mutant will win the competition.

Second, we describe the effects of genetic drift~(demographic fluctuations), which can lead to the extinction of a beneficial mutant in both spatial and well-mixed populations~\cite{gillespie:book}. In well-mixed populations, the establishment or fixation probability of a mutant depends only on its fitness advantage and the strength of genetic drift. In expanding populations, the location of the mutant also affects its ultimate fate. For example, if the initial conditions are chosen to be orthogonal to the eigenfunction with the largest eigenvalue, then the mutant will not grow at the rate given by the largest eigenvalue. Generic initial conditions, including the ones due to a spontaneous mutation, however, have a nonzero projection on the leading eigenfunction, and the mutant will eventually grow at a rate given by the largest eigenvalue, even if the net population density may slightly decline initially. The magnitude of this projection determines the time that the mutant spends at low densities and therefore experiences strong fluctuations, which can drive it to extinction. As a result, the fixation probability of a mutant depends not only on the magnitude of~$\lambda$, but also on the point of origin. Mutants that occur near the maximum of the leading eigenfunction have a larger fixation probability compared to mutants occurring away from this maximum, for example well at the back of the front. This is a well-known issue in spatial populations and is not specific to our analysis~\cite{hallatschek:diversity,lehe:surfing}. The net effect on species evolution is that only mutations occurring within a certain spatial region are contributing to the adaptation. This is a reassuring conclusions; otherwise, the adaptation of large, spatially extended populations would be extremely rapid.

\textit{Symmetry of invasion fitnesses}\\
The rate at which the mutant type grows in the background of the wild type is known as the invasion fitness in the field of adaptive dynamics~\cite{waxman:review_adaptive_dynamics}. Here, we discuss the properties of invasion fitness under the exchange of the types: type two invading type one instead of type one invading type two. Concretely, for all values of~$\rho^*$ and~$p\approx1$, our exact solution shows that the fitness advantage of type one in the background of type two equals the fitness disadvantage of type two in the background of type one. However, when the dispersal rates are very unequal, we observe an asymmetry under the exchange of~$D_1$ and~$D_2$. The reason for this asymmetry is that the fitness is the property of both an organism and its environment, which, in this case, is the presence of a competing type. The interaction between the organism and its environment is in general nonlinear and results in asymmetric rates of invasion. Thus, our results illustrate that symmetric invasion fitnesses do not fully capture the complexity of phenotype evolution.

\textit{Robustness of results to model assumptions}\\
In range expansions, one typically distinguishes between short-range and long-range dispersal kernels. Short-range kernels are described well by reaction-diffusion equations at long spatial and temporal scales due to the central limit theorem. Indeed, our numerical simulations with discrete jumps between nearest neighbors give the same results as the analytic solution of the continuous equations. The effects of long-range dispersal are more subtle, and, fortunately, there are few invaders with the capabilities to travel long range. For so called pulled expansions, which occur when the Allee effect is absent or weak, long-range dispersal can lead to expansion acceleration over time~\cite{kot:csdt}. Although such dynamics are not captured by our analysis, it is important to emphasize that they lead to the same evolutionary outcome as we described in this paper. Our analysis predicts that faster dispersers would be favored in this regime~(Fig.~\ref{fig:phase_diagram}). The same dynamics are expected for species with long-range dispersal because organisms that land far away from the ancestral population will be able to start a new invasion and thus colonize all the areas ahead of the expansion front. When the Allee effect is strong, long-range dispersal does not lead to qualitatively new dynamics like wave acceleration because the organisms that disperse too far find themselves at densities below the Allee threshold and, therefore, fail to establish and start a new expansion~\cite{kot:csdt}. Because these long-range dispersal events effectively lead to death, mutants that increase their dispersal rate will die more frequently and would not be selected, similar to the conclusion of our analysis. In summary, different dispersal kernels should not lead to qualitatively new results. The quantitative results will of course be different since they depend on the model details including the type of dispersal and the form of~$g(c)$.

%\clearpage
%**************************************************************************
% Simulations
%**************************************************************************
\textbf{\large Simulations}

\textit{Numerical solution of equations~(\ref{eq:model})}\\
We solved equations~(\ref{eq:model}) using an explicit finite difference method with the following discretization:

\begin{equation}
	\begin{aligned}
		& c_{1}(t+\Delta t,x) = c_{1}(t,x) + D_{1}\frac{c_{1}(t,x+\Delta x) + c_{1}(t, x-\Delta x) -2c_{1}(t,x)}{\Delta x^2}\Delta t + c_{1}g(c)\Delta t,\\
		& c_{2}(t+\Delta t,x) = c_{2}(t,x) + D_{2}\frac{c_{2}(t,x+\Delta x) + c_{2}(t, x-\Delta x) -2c_{2}(t,x)}{\Delta x^2}\Delta t + c_{2}g(c)\Delta t.
	\end{aligned}
	\label{eq:model_discretized}
\end{equation}

\noindent Equations~(\ref{eq:model_discretized}) were iterated in a spatial domain of length at least~$35$ with at least~$700$ discretization points. We kept~$D_2=1$,~$g=1$,~$K=1$ constant and varied~$D_1$ and~$c^*$. The temporal discretization~$\Delta t$ was set to~$0.01\Delta x^2$ to ensure both accuracy and stability of the numerical algorithm. The temporal duration of the simulation was varied with~$\Delta D$ to ensure that sufficient data are available to estimate the fitness difference between the strains. The simulations were started with the left half of the habitat occupied by the species with~$c=K$. The fraction of the first type was set to~$10^{-2}$ or~$10^{-3}$ uniformly in space. As the expansion approached the right side of the simulation domain, we shifted the simulation domain to recenter the population. 

For each simulation, we confirmed that the expansion velocities agreed with equation~(\ref{eq:profile}) within 1\% error. Then, the fraction of the first mutant~$f(t)$ was estimated as the average~$f(t,x)$ across all the discrete points where the solution was computed. To obtain the selective advantage of the first species~$\lambda$, we fitted~$\ln(f(t)/(1-f(t))$ to~$\lambda t+\mathrm{const}$.

\textit{Stochastic simulations}\\
Stochastic simulations were implemented as the stepping-stone model~\cite{kimura:ssm} and Levins' metapopulation model~\cite{levins:model} with discrete generations, which relaxed the continuity assumption of equations~(\ref{eq:model}) and allowed for landscape fragmentation. Each generation consisted of a growth and dispersal phases. The dispersal phase, allowed each organism to migrate to one of the two nearby patches with identical probability equal to~$m/2$. The growth phase was implemented as a birth-death process with the per capita birth rate equal to~$rc/K + rcc^*/K^2$ and the per capita death rate equal to~$rc^2/K^2 + rc^*/K$, which lead to the dynamics described by equation~(\ref{eq:growth}) in the continuous limit. The dispersal rate of each newly-born organism could be mutated with probability~$\mu=10^{-4}$ to either increase or decrease by~10\%. We capped the maximal dispersal rate at~$m=0.5$. Similar to the deterministic simulations, we re-centered the simulation box when necessary. The number of patches was~150 and the carrying capacity was~$10^4$.

\bibliography{references}
\end{document}